\documentclass[twocolumn,showpacs,preprintnumbers,amsmath,amssymb]{revtex4-1}

\usepackage{graphicx}
\usepackage{dcolumn}
\usepackage{bm}
\usepackage{amssymb}
\usepackage{multirow}
\usepackage{braket}
\usepackage{amsmath}
 \usepackage[german,english]{babel}
\makeatletter
\newcommand{\vast}{\bBigg@{4}}
\newcommand{\Vast}{\bBigg@{5}}
\makeatother

\begin{document}

\title{Photoionization microscopy in terms of local frame transformation theory}

\author{P. Giannakeas}
\email{pgiannak@purdue.edu}
\affiliation{Department of Physics and Astronomy, Purdue University, West Lafayette, Indiana 47907, USA}

\author{F. Robicheaux}
\email{robichf@purdue.edu}
\affiliation{Department of Physics and Astronomy, Purdue University, West Lafayette, Indiana 47907, USA}

\author{Chris H. Greene}
\email{chgreene@purdue.edu}
\affiliation{Department of Physics and Astronomy, Purdue University, West Lafayette, Indiana 47907, USA}

\date{\today}

\pacs{32.80.Fb, 32.60.+i, 07.81.+a}

\begin{abstract}
Two-photon ionization of an alkali-metal atom in the presence of a uniform electric field is investigated using a standardized form of local frame transformation and generalized quantum defect theory. 
The relevant long-range quantum defect parameters in the combined Coulombic plus Stark potential is calculated with eigenchannel R-matrix theory applied in the downstream parabolic coordinate $\eta$.
The present formulation permits us to express the corresponding microscopy observables in terms of the local frame transformation, and it gives a critical test of the accuracy of the Harmin-Fano theory permitting a scholastic investigation of the claims presented in Zhao {\it et al.} [Phys. Rev. A 86, 053413 (2012)].
\end{abstract}

\maketitle
\section{Introduction}
The photoabsorption spectrum of an alkali-metal atom in the presence of a uniform electric field constitutes a fundamental testbed for atomic physics. Through the past few decades, study of this class of systems has provided key insights into their structure and chemical properties, as well as the nonperturbative effect of an applied external field. 
The response of the lower energy eigenstates of any alkali-metal atom to a laboratory strength electric field is perturbative and can be described in terms of the static atomic polarizability. 
For states high in the Rydberg series or in the ionization continuum, however, even a modest field strength nonperturbatively modifies the nature of the energy eigenstates.  

In fact this problem touches on fundamental issues concerning the description of nonseparable quantum mechanical systems. 
The Stark effect of alkali-metal atoms is one of the simpler prototypes of such systems, because the short-distance electron motion is nearly separable in spherical coordinates while the intermediate- and long-distance motion is almost exactly separable in parabolic coordinates. 
The evolution of a quantum electron wave function from small to large distances thus involves a transformation, termed a {\it local} frame transformation (LFT) because it is derived in a localized region of space. 
(The extent of this region is typically limited to within 10-20 a.u. between the electron and the nucleus.)

When one encounters a problem of nonrelativistic quantum mechanics where the Schr\"odinger equation is nonseparable, one usually anticipates that the system will require a complicated numerical treatment. 
This is the first and most common approach even if the nonseparability is limited to only two coordinates as is the case with the nonhydrogenic Stark effect since the azimuthal angle $\phi$ is separable for this problem (aside from the comparatively weak spin-orbit coupling). 
Thus it was a major breakthrough when papers by Fano \cite{fano1981stark} and Harmin \cite{harminpra1982,harminprl1982,harmin1981hydrogenic} showed in the early 1980s how the problem can be solved analytically and almost completely using ideas based on the frame transformation theory and quantum defect theory. 
Since that body of work introduced the LFT method, it has been generalized to other systems that are similar in having an intermediate region of space where the wave equation is separable in both the small- and large-distance coordinate systems. 
Example applications include diverse systems such as negative ion photodetachment in either an external magnetic \cite{Greene1987pra} or electric field \cite{WongRauGreene1988pra,RauWong1988pra,GreeneRouze1988ZPhys,SlonimGreene1991}, and confinement-induced resonances in ultracold atom-atom scattering \cite{GrangerBlume2004prl,Giannakeas2012pra,hess2014pra,Zhang2013pra} or dipole-dipole collisions \cite{giannakeas2013prl}.

The LFT theory has been demonstrated by now to have great effectiveness in reproducing experimental spectra and collision properties as well as accurate theoretical results derived using other methods including ``brute force'' computations \cite{stevens1996precision}.
The deviations between highly accurate R-matrix calculations and the LFT method were found in Ref. \cite{stevens1996precision} to be around 0.1\% for resonance positions in the $^7\rm{Li}$ Stark effect. 
The LFT is evolving as a general tool that can solve this class of nonseparable quantum mechanical problems, but it must be kept in mind that it is an approximate theory.
It is therefore desirable to quantify the approximations made, in order to understand its regimes of applicability and where it is likely to fail.  

The goal of the present study is to provide a critical assessment of the accuracy of the LFT, concentrating in particular on observables related to photoionization microscopy. 
The experiments in this field \cite{cohen13,itatani2004tomographic,bordas03,Nicole02} have focused on the theoretical proposal that the probability distribution of an ejected slow continuum electron can be measured on a position-sensitive detector at a large distance from the nucleus \cite{ost1,ost2,ost3,ost4}.

While the Harmin-Fano LFT theory has been shown in the 1980s and 1990s to describe the total photoabsorption Stark spectra in one-electron \cite{harminpra1982,harminprl1982,stevens1996precision} and two-electron \cite{Armstrong1993prl,Armstrong1994pra,Robicheaux1999pra} Rydberg states, examination of a differential observable such as the photodetachment \cite{Blondel1996prl} or photoionization \cite{Texier2005pra} microscopy probability distribution should in principle yield a sharper test of the LFT. 
Indeed, a recent study by Zhao, Fabrikant, Du, and Bordas \cite{zhao12} identifies noticeable discrepancies between Harmin's LFT Stark effect theory and presumably more accurate coupled-channel calculations. 
Particularly in view of the extended applications of LFT theory to diverse physical contexts, such as the confinement-induced resonance systems noted above, a deeper understanding of the strengths and limitations of the LFT is desirable.

In this paper we employ R-matrix theory in a fully quantal implementation of the Harmin local frame transformation, instead of relying on semiclassical wave mechanics as he did in Refs.\cite{ harminpra1982,harminprl1982,harmin1981hydrogenic}. 
This allows us to disentangle errors associated with the WKB approximation from those deriving from the LFT approximation itself.
For the most part this causes only small differences from the original WKB treatment consistent with Ref. \cite{stevens1996precision}, but it is occasionally significant, for instance for the resonant states located very close to the top of the Stark barrier. 
Another goal of this study is to standardize the local frame transformation theory to fully specify the asymptotic form of the wave function which is needed to describe other observables such as the spatial distribution function (differential cross section) that is measured in photoionization microscopy.

We also revisit the interconnection of the irregular solutions from spherical to parabolic coordinates through the matching of the spherical and parabolic Green's functions in the small distance range where the electric field is far weaker than the Coulomb interaction. 
This allows us to re-examine the way the irregular solutions are specified in the Fano-Harmin LFT, which is at the heart of the LFT method but one of the main focal points of criticism leveled by Zhao {\it et al.} \cite{zhao12}.

Because Zhao {\it et al.} \cite{zhao12} raise serious criticisms of the LFT theory, it is important to further test their claims of error and their interpretation of the sources of error.  Their contentions can be summarized as follows:

{\it (i)}  The Harmin-Fano LFT quite accurately describes the total photoionization cross section, but it has significant errors in its prediction of the differential cross section that would be measured in a photoionization microscopy experiment. 
This is deduced by comparing the results from the approximate LFT with a numerical calculation that those authors regard as essentially exact.

{\it (ii)}  The errors are greatest when the atomic quantum defects are large, and almost negligible for an atom like hydrogen which has vanishing quantum defects. 
They then present evidence that they have identified the source of those errors in the LFT theory, namely the procedure first identified by Fano that predicts how the irregular spherical solution evolves at large distances into parabolic coordinate solutions.
Their calculations are claimed to suggest that the local frame transformation of the solution regular at the origin from spherical to parabolic coordinates is correctly described by the LFT, but the irregular solution transformation is incorrect.

One of our major conclusions from our exploration of the Ref.\cite{ zhao12} claimed problems with the Harmin-Fano LFT is that both claims are erroneous; their incorrect conclusions apparently resulted from their insufficient attention to detail in their numerical calculations.
Specifically, our calculations for the photoionization microscopy of Na atoms ionized via a two photon process in $\pi$ polarized laser fields do not exhibit the large and qualitative inaccuracies which were mentioned in Ref.\cite{zhao12}; for the same cases studied by Zhao {\it et al.}, we obtain excellent agreement between the approximate LFT theory and our virtually exact numerical calculations. 
Nevertheless some minor discrepancies are noted which may indicate minor inaccuracies of the local frame transformation theory.

This paper is organized as follows: Section II focuses on the local frame transformation theory of the Stark effect and present a general discussion of the physical content of the theory, including a description of the relevant mappings of the regular and irregular solutions of the Coulomb and Stark-Coulomb Schr\"odinger equation.
Section III reformulates the local frame transformation theory properly, including a description of the asymptotic electron wave function.
In addition, this Section defines all of the relevant scattering observables.
Section IV discusses a numerical implementation based on a two-surface implementation of the eigenchannel R-matrix theory.
This toolkit permits us to perform accurate quantal calculations in terms of the local frame transformation theory, without relying on the semiclassical wave mechanics adopted in Harmin's implementation.
Section V is devoted to discussion of our recent finding in comparison with the conclusions of Ref.\cite{zhao12}.
Finally, Section VI summarizes and concludes our analysis.

\section{Local frame transformation theory of the Stark effect}
This section reviews the local frame transformation theory (LFT) for the non-hydrogenic Stark effect, utilizing the same nomenclature introduced by Harmin \cite{harminpra1982,harminprl1982,harmin1981hydrogenic}.
The crucial parts of the corresponding theory are highlighted developing its standardized formulation. 

\subsection{General considerations}

In the case of alkali-metal atoms at small length scales the impact of the alkali-metal ion core on the motion of the valence electron outside the core can be described effectively by a phase-shifted radial wave function:
\begin{equation}
 \Psi_{\epsilon \ell m}(\mathbf{r})=\frac{1}{r}Y_{\ell m}(\theta, \phi)\big[ f_{\epsilon \ell}(r) \cos \delta_\ell-g_{\epsilon \ell}(r) \sin \delta_\ell \big],~~r>r_0,
\label{spherwave}
 \end{equation}
where the $Y_{\ell m}(\theta, \phi)$ are the spherical harmonic functions of orbital angular momentum $\ell$ and projection $m$. 
$r_0$ indicates the effective radius of the core, $\delta_\ell$ denotes the phase that the electron acquires due to the alkali-metal ion core.
These phases are associated with the quantum defect parameters, $\mu_\ell$, according to the relation $\delta_\ell=\pi \mu_\ell$.
The pair of $\{f,g\}$ wave functions designate the regular and irregular Coulomb ones respectively whose Wronskian is $W[f,g]=2/\pi$.
We remark that this effective radius $r_0$ is placed close to the origin where the Coulomb field prevails over the external electric field. 
Therefore, the effect on the phases $\delta_\ell$ from the external field can be neglected.
Note that atomic units are employed everywhere, otherwise is explicitly stated.

At distances $r\gg r_0$ the outermost electron of the non-hydorgenic atom is in the presence of a homogeneous static electric field oriented in the $z$-direction. 
The separability of the center-of-mass and relative degrees of freedom permits us to describe all the relevant physics by the following Schr\"odinger equation in the relative frame of reference:
\begin{equation}
 \bigg(-\frac{1}{2}\nabla^2-\frac{1}{r}+ F z-\epsilon \bigg)\psi(\mathbf{r})=0,
 \label{eq1}
\end{equation}
where $F$ indicates the strength of the electric field, $r$ corresponds to the interparticle distance and $\epsilon$ is the total colliding energy.
Note that Eq.~(\ref{eq1}) is invariant under rotations around the polarization axis, namely the corresponding azimuthal quantum number $m$ is a good one.
In contrast, the total orbital angular momentum is not conserved, which shows up as a coupling among different $\ell$ states.
The latter challenge, however, can be circumvented by employing a coordinate transformation which results in a fully separable Schr\"odinger equation.
Hence, in parabolic coordinates $\xi=r+z$, $\eta=r-z$ and $\phi=\tan^{-1}(x/y)$, Eq.~(\ref{eq1}) reads:
\begin{equation}
 \frac{d^2}{d\xi^2}\Xi_{\beta m}^{ \epsilon F}(\xi)+\bigg(\frac{\epsilon}{2}+\frac{1-m^2}{4\xi^2}+\frac{\beta}{\xi}-\frac{F}{4}\xi\bigg)\Xi_{\beta m}^{ \epsilon F}(\xi)=0,
 \label{eq2}
\end{equation}

\begin{equation}
 \frac{d^2}{d\eta^2}\Upsilon_{\beta m}^{ \epsilon F}(\eta)+\bigg(\frac{\epsilon}{2}+\frac{1-m^2}{4\eta^2}+\frac{1-\beta}{\eta}+\frac{F}{4} \eta\bigg)\Upsilon_{\beta m}^{ \epsilon F} (\eta)=0,
 \label{eq3}
\end{equation}
where $\beta$ is the {\it effective charge} and $\epsilon$, $F$ are the energy and the field strength in atomic units.
We remark that Eq.~(\ref{eq2}) in the $\xi$ degrees of freedom describes the bounded motion of the electron since as $\xi \to \infty$ the term with the electric field steadily increases.
This means that the $\Xi$ wave function vanishes as $\xi \to \infty$ for every energy $\epsilon$ at particular values of the effective charge $\beta$.
Thus, Eq.~(\ref{eq2}) can be regarded as a generalized eigenvalue equation where for each quantized $\beta \equiv \beta_{n_1}$ the $\Xi_{\beta m}^{\epsilon F} \equiv \Xi_{n_1 m}^{\epsilon F}$ wave function possesses $n_1$ nodes. 
In this case the wave functions $\Xi_{n_1 m}^{\epsilon F}(\xi)$ possess the following properties:
\begin{itemize}
 \item  Near the origin $\Xi_{n_1 m}^{\epsilon F}$ behaves as: $\Xi_{n_1 m}^{\epsilon F}(\xi\to 0)\sim N^F_{\xi} \xi^\frac{m+1}{2}[1+O(\xi)]$, where $N^F_\xi$ is an energy-field dependent amplitude and must be determined numerically in general.
 \item The wave function $\Xi_{n_1 m}^{\epsilon F}$ obeys the following normalization condition: $\int_0^\infty \frac{[\Xi_{n_1 m}^{\epsilon F}(\xi)]^2}{\xi}\rm{d} \xi=1.$
\end{itemize}

On the other hand Eq.~(\ref{eq3}) describes solely the motion of the electron in the $\eta$ degree of freedom which is unbounded.
As $\eta \to \infty$ the term with the electric field steadily decreases which in combination with the coulomb potential forms a barrier that often has a local maximum.
Hence, for specific values of energy, field strength and effective charge the corresponding wave function $\Upsilon_{\beta m}^{\epsilon F} \equiv \Upsilon_{n_1 m}^{\epsilon F}$ propagates either above or below the barrier local maximum where the states $n_1$ define asymptotic channels for the scattering wave function in the $\eta$ degrees of freedom.
Note that for $\beta_{n_1}>1$, the Coulomb term in Eq.~(\ref{eq3}) becomes repulsive and therefore no barrier formation occurs.
Since Eq.~(\ref{eq3}) is associated with the unbounded motion of the electron it possesses two solutions, namely the regular $\Upsilon_{n_1 m}^{\epsilon F}(\eta)$ and the irregular ones $\bar{\Upsilon}_{n_1 m}^{\epsilon F}(\eta)$.
This set of solutions has the following properties:
\begin{itemize}
 \item Close to the origin and before the barrier the irregular solutions $\bar{\Upsilon}_{n_1 m}^{\epsilon F}(\eta)$ lag by $\pi/2$ the regular ones, namely $\Upsilon_{n_1 m}^{\epsilon F}(\eta)$. Note that their normalization follows Harmin's definition \cite{harminpra1982} and is clarified below.
 \item Near the origin the regular solutions vanish according to the relation:  $\Upsilon_{n_1 m}^{\epsilon F}(\eta \to 0)\sim N^F_{\eta} \eta^\frac{m+1}{2}[1+O(\eta)]$, where $N^F_\eta$ is an energy- and field-dependent amplitude and must be determined numerically in general.
\end{itemize}

Let us now specify the behavior of the pair solutions $\{\Upsilon_{n_1 m}^{\epsilon F},\bar{\Upsilon}_{n_1 m}^{\epsilon F}\}$ at distances after the barrier. 
Indeed, the regular and irregular functions can be written in the following WKB form:
\small
\begin{eqnarray}
&~&\Upsilon_{n_1 m}^{\epsilon F}(\eta \gg \eta_0) \to \sqrt{\frac{2}{\pi k(\eta)}} \sin \bigg[\int^{\eta}_{\eta_0}k(\eta')d\eta'+\frac{\pi}{4}+\delta_{n_1}\bigg]
\label{regU}
\end{eqnarray}
\begin{eqnarray}
&~&\bar{\Upsilon}_{n_1 m}^{\epsilon F}(\eta \gg \eta_0) \to \sqrt{\frac{2}{\pi k(\eta)}} \sin \bigg[\int^{\eta}_{\eta_0}k(\eta')d\eta'+\frac{\pi}{4}+\delta_{n_1}- \gamma_{n_1}\bigg],
\label{irregU}
\end{eqnarray}
\normalsize
where $k(\eta)=\sqrt{-m^2/\eta^2+(1-\beta_{n_1})/\eta+\epsilon/2+F\eta/4}$ is the local momentum term with the Langer correction being included, $\eta_0$ is the position of the outermost classical turning point and the phase $\delta_{n_1}$ is the absolute phase induced by the combined Coulomb and electric fields.
The phase $\gamma_{n_1}$ corresponds to the relative phase between the regular and irregular functions, namely $\{\Upsilon, \bar{\Upsilon}\}$. 
We recall that at short distances their relative phase is exactly $\pi/2$, though as they probe the barrier at larger distances their relative phase is altered and hence after the barrier the {\it short range} regular and irregular functions differ by $0<\gamma_{n_1}<\pi$ and not just $\pi/2$.
We should remark that after the barrier the amplitudes of the pair $\{\Upsilon, \bar{\Upsilon}\}$ are equal to each other and their relative phase in general differs from $\pi/2$.
On the other hand, at shorter distances before the barrier the amplitudes of the  $\{\Upsilon, \bar{\Upsilon}\}$ basically are not equal to each other and their relative phase is exactly $\pi/2$.
This ensures that the Wronskian of the corresponding solutions possesses the same value at all distances and provides us with insight into the interconnection between amplitudes and relative phases.

The key concept of Harmin's theoretical framework is to associate the relevant phases at short distances in the absence of an external field, i.e. $\delta_\ell$ (see Eq.~(\ref{spherwave})) to the scattering phases at large distances where the electric field contributions cannot be neglected. 
This can be achieved by mapping the corresponding regular and irregular solutions from spherical to parabolic-cylindrical coordinates as we discuss in the following.

\subsection{Mapping of the regular functions from spherical to parabolic-cylindrical coordinates}

The most intuitive aspect embedded in the present problem is that the Hamiltonian of the motion of the electron right outside the core possesses a spherical symmetry which in turn at greater distances due to the field becomes parabolic-cylindrically symmetric.
Therefore, a proper coordinate transformation of the corresponding {\it energy normalized} wave functions from spherical to parabolic cylindrical coordinates will permit us to {\it propagate} to asymptotic distances the relevant scattering or photoionization events initiated near the core. 
Indeed at distances $r\ll F^{-1/2}$ the regular functions in spherical coordinates are related to the parabolic cylindrical ones according to the following relation:

\begin{eqnarray}
 \psi_{n_1 m}^{\epsilon F}(\mathbf{r})&=&\frac{e^{i m \phi}}{\sqrt{2 \pi}}\frac{\Xi_{n_1 m}^{\epsilon F}(\xi)}{\sqrt{\xi}}\frac{\Upsilon_{n_1 m}^{\epsilon F}(\eta)}{\sqrt{\eta}}\nonumber \\
 &=&\sum_\ell U_{n_1 \ell}^{\epsilon Fm} \frac{f_{\epsilon \ell m}(\mathbf{r})}{r}, ~~\rm{for}~~ r \ll F^{-1/2},
\label{eq4}
\end{eqnarray}
where $f_{\epsilon \ell m}(\mathbf{r})$ are the regular solutions in spherical coordinates with $\ell$ being the orbital angular momentum quantum number.
The small distance behavior is $f_{\epsilon \ell m}(\mathbf{r})\approx N_{\epsilon \ell}Y_{\ell m}(\theta,\phi)r^{\ell+1}[1+O(r)]$ with $N_{\epsilon \ell}$ a normalization constant (see Eq.~(13) in Ref. \cite{harminpra1982}).
Therefore, from the behavior at small distances of the parabolic-cylindrical and spherical solutions the frame transformation $U_{n_1 \ell}^{\epsilon Fm}$ has the following form:
\begin{widetext}
\begin{equation}
 U_{n_1 \ell}^{\epsilon F m}=\frac{N^F_\xi N^F_\eta}{N_{\epsilon \ell}}\frac{(-1)^m\sqrt{4 \ell+2}m!^2}{(2 \ell+1)!! \sqrt{(\ell+m)!(\ell-m)!}}\sum_{k}^{\ell-m} (-1)^k  \binom{\ell-m}{k}\binom{\ell+m}{\ell-k} 
 \frac{\nu^{m-\ell}\Gamma(n_1+1)\Gamma(\nu-n_1-m)}{\Gamma(n_1+1-k)\Gamma(\nu-n_1+k-\ell)},
 \label{eq5}
\end{equation}
\end{widetext}
where $n_1=\beta_{n_1} \nu -1/2-m/2$ and $\nu=1/\sqrt{-2 \epsilon}$.

\begin{figure}[t!]
\includegraphics[scale=0.75]{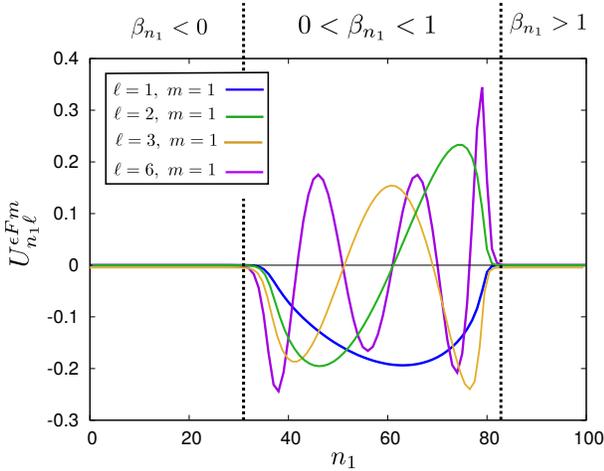}
\caption{(color online). The matrix elements of the local frame transformation $U_{n_1 \ell}^{\epsilon Fm}$ versus the number of states $n_1$ for $m=1$ where the angular momentum acquires the values  $\ell=1,2,3~\rm{and}~6$. The electric field strength is $F= 640$ V/cm and total collisional energy is $\epsilon=135.8231$ cm$^{-1}$. The vertical dashed lines indicate the sign and the interval of values of the $\beta_{n_1}$.}
\label{fig1}
\end{figure}

Fig. \ref{fig1} plots the elements of the local frame transformation $U$ in Eq.~(\ref{eq5}) as functions of the number of states $n_1$, where again the integers $n_1$ label the eigenvalues $\beta_{n_{1}}$.
The local frame transformation $U$ is plotted for four different angular momenta, namely $\ell=1,2,3~\rm{and}~6$ where we set $m=1$ at energy $\epsilon=135.8231$ cm$^{-1}$ and field $F= 640$ V/cm.
One sees that the local frame transformation $U$ becomes significant in the interval $n_1 \in (38,79)$ which essentially corresponds to $\beta_{n_1} \in(0,1)$. 
For $\beta_{n_1}<0$ or $\beta_{n_1}>1$ the local frame transformation vanishes rapidly.
This behavior mainly arises from the normalization amplitudes $N_\xi^F$ and $N_\eta^F$, which obey the following relations:
\begin{equation}
 N_\xi^F \sim \frac{\beta_{n_1}}{1-e^{-2 \pi \beta_{n_1}/k}}~~\rm{and}~~  N_\eta^F \sim \frac{(1-\beta_{n_1})}{1-e^{-2 \pi(1- \beta_{n_1})/k}}.
 \label{eq6}
\end{equation}
Note that these expressions are approximately valid only for positive energies and they are exact for $F=0$.

From the expressions in Eq.~(\ref{eq6}) it becomes evident that for negative eigenvalues $\beta_{n_1}$ the amplitude $N_\xi^F$ vanishes exponentially while $N_\eta^F$ remains practically finite. 
Similarly, for the case of $\beta_{n_1}>1$ the amplitude $N_\eta^F$ vanishes exponentially, and these result in the behavior depicted in Fig.\ref{fig1}.
Another aspect of the local frame transformation $U$ is its nodal pattern shown in Fig.\ref{fig1}.
For increasing $\ell$ the corresponding number of nodes increases as well.
For $m=1$, every $U_{n_1 \ell}^{\epsilon Fm}$ possesses $\ell-1$ nodes.

\subsection{Mapping of the irregular functions from spherical to parabolic-cylindrical coordinates}

Having established the mapping between the regular solutions of the wave function in spherical and parabolic-cylindrical coordinates, the following focuses on the relation between the irregular ones.

The irregular solution in the parabolic-cylindrical coordinates has the following form:
\begin{equation}
\chi_{n_1 m}^{\epsilon F}(\mathbf{r})=\frac{e^{i m \phi}}{\sqrt{2 \pi}}\frac{\Xi_{n_1 m}^{\epsilon F}(\xi)}{\sqrt{\xi}}\frac{\bar{\Upsilon}_{n_1 m}^{\epsilon F}(\eta)}{\sqrt{\eta}},
\label{eq7}
\end{equation}
Recall that 
In order to relate Eq.~(\ref{eq7}) to the irregular functions in spherical coordinates we employ Green's functions as was initially suggested in \cite{fano1981stark}.
More specifically, the principal value Green's function for the pure Coulomb Hamiltonian $G_P^{(C)}(\mathbf{r,r'})$, is matched with a Green's function of the Coulomb plus Stark Hamiltonian $G^{(C+F)}(\mathbf{r,r'})$, which is expressed in parabolic-cylindrical coordinates.

Of course, in general the two Green's functions differ from each other since they correspond to different Schr\"odinger equations.
However, at small distances the field term in the Stark Hamiltonian becomes negligible in comparison with the Coulomb term.
Therefore, in this restricted region of the configuration space, i.e. $r \ll F^{-1/2}$, the Stark Hamiltonian is virtually identical to the Coulomb Hamiltonian, whereby the corresponding Green's functions are equivalent to an excellent approximation.
We refer to this region as the Coulomb zone.

For positive energies recall that the principal value Green's function is uniquely defined in the infinite configuration space, and it consists of a sum of products of regular and irregular functions. 
The employed regular and irregular functions are defined such that their relative phase is exactly $\pi/2$ asymptotically \cite{Rodberg1970,Economou200606}.
Therefore, according to the above mentioned arguments the principal value Green's function obeys the relation expressed in spherical coordinates:
\small
\begin{equation}
 G^{(C)}_P(\mathbf{r,r'})= \frac{\pi}{r r'} \sum_{\ell, m} f_{\epsilon \ell m}(\mathbf{r})g_{\epsilon \ell m}(\mathbf{r'}),~~r<r'
 \label{eq8}
\end{equation}
\normalsize
where the $\{f,g\}$ solutions correspond to the regular and irregular functions as they are defined in Eq.~(\ref{spherwave})
Note that the principal value Green's function of the Coulomb Hamiltonian in spherical and in parabolic-cylindrical coordinates are equal to each other, namely $G_P^{(C),\rm{sc}}\equiv G_P^{(C),\rm{pcc}}$ (the abbreviations sc and pcc stand for spherical and parabolic-cylindrical coordinates, respectively).

On the other hand for negative energies, by analytically continuing the $\{f,g\}$ Coulombic functions across the threshold yields the relation $\mathcal{G}^{(C),\rm{sc}}\equiv \mathcal{G}^{(C),\rm{pcc}}$ .
The $\mathcal{G}^{(C)}$ is the so called {\it smooth} Green's function which is related to a Green's function bounded at $r=0$ and at infinity according to the expression \cite{greene1979general}:
\begin{equation}
 G^{(C)}(\mathbf{r,r'})=\mathcal{G}^{(C)}(\mathbf{r,r'})+\frac{\pi}{r r'} \sum_\ell f_{\epsilon \ell m}(\mathbf{r}) \cot \beta(\epsilon) f_{\epsilon \ell m}(\mathbf{r'}),
\label{smoothg}
\end{equation}
where $\beta(\epsilon)=\pi (\nu-\ell)$ with $\nu=1/\sqrt{-2\epsilon}$ is the phase accumulated from $r=0$ up to $r\to \infty$.
Assume that $\epsilon_n$ (i.e. $\nu \to n \in \aleph^*$) are the eigenergies specified by imposing the boundary condition at infinity where $n$ denotes a counting index of the corresponding bound states.
Then in the right hand side of Eq.~(\ref{smoothg}) the second term at energies $\epsilon=\epsilon_n$ diverges while the first term is free of poles.
The smooth Green's function is identified as the one where the two linearly independent solutions have their relative phase equal to $\pi/2$ {\it at small distances}. 
Furthermore, the singularities in Eq.~(\ref{smoothg}) originate from imposing the boundary condition at infinity, though in the spirit of multichannel quantum defect theory we can drop this consideration and solely employ the $\mathcal{G}^{(C)}$ which in spherical coordinates reads
\small
\begin{equation}
 \mathcal{G}^{(C)}(\mathbf{r,r'})= \frac{\pi}{r r'} \sum_{\ell, m} f_{\epsilon \ell m}(\mathbf{r})g_{\epsilon \ell m}(\mathbf{r'}),~~r<r'~~\rm{for}~\epsilon<0.
 \label{smoothspher}
\end{equation}
\normalsize

In view of the now established equality between the principal value (smooth) Green's functions at positive (negative) energies in spherical and parabolic cylindrical coordinates for the pure Coulomb Hamiltonian, the discussion can proceed to the Stark Hamiltonian.
Hence as mentioned above in the Coulomb zone, i.e. $r \ll F^{-1/2}$, the Stark Hamiltonian is approximately equal to the pure Coulomb one.
This implies the existence of a Green's function, $G^{(C+F)}$, for the Stark Hamiltonian which is equal to the $G_P^{(C),\rm{pcc}}$ ($\mathcal{G}^{(C),\rm{pcc}}$), and which in turn is equal to Eq.~(\ref{eq8}) [Eq.~(\ref{smoothspher})] at positive (negative) energies.
More specifically, the $G^{(C+F)}$ the Green's function expressed in parabolic-cylindrical coordinates is given by the expression:
\small
\begin{equation}
 G^{(C+F)}(\mathbf{r,r'})= 2 \sum_{n_1, m} \frac{\psi^{\epsilon F}_{n_1 m}(\mathbf{r})\chi^{\epsilon F}_{n_1 m}(\mathbf{r'})}{W(\Upsilon_{n_1 m}^{\epsilon F},\bar{\Upsilon}_{n_1 m}^{\epsilon F})},~\rm{for}~\eta< \eta' \ll F^{-1/2},
 \label{eq9}
\end{equation}
\normalsize
where the functions $\{\psi, \chi\}$ are the regular and irregular solutions of the Stark Hamiltonian, which at small distances (in the classically allowed region) have a relative phase of $\pi/2$. 
This originates from $\pi/2$ relative phase of the $\{\Upsilon, \bar{\Upsilon}\}$ as was mentioned is subsection A.
The Wronskian $W[\Upsilon_{n_1 m}^{\epsilon F},\bar{\Upsilon}_{n_1 m}^{\epsilon F}]=(2/\pi) \sin \gamma_{n_1}$ yields $\{\psi, \chi\}$ solutions have  the same energy normalization as in the $\{f, g\}$ coulomb functions.

We should point out that Eq.~(\ref{eq9}) is not the principal value Green's function of the Stark Hamiltonian.
Indeed, it can be shown that principal value Green's function of the Stark Hamiltonian, namely $G^{(C+F)}_P$ and the Green's function $G^{(C+F)}$ obey the following relation:
\small
\begin{eqnarray}
  G^{(C+F)}(\mathbf{r,r'})&=&G^{(C+F)}_P(\mathbf{r,r'})\nonumber \\
  &+&\sum_{n_1}\cot \gamma_{n_1} \psi^{\epsilon F}_{n_1 m}(\mathbf{r})\psi^{\epsilon F}_{n_1 m}(\mathbf{r'}),
  \label{greenpv}
\end{eqnarray}
\normalsize
where we observe that either for positive energies or for $n_1$ channels which lie above the saddle point of the Stark barrier the second term vanishes.
This occurs due to the fact that $\gamma_{n_1}\approx \pi/2$ since the barrier does not alter the relative phases between the regular and irregular solutions.
For the cases where the barrier effects are absent the $ G^{(C+F)}$ is the principal value Green's function of the Stark Hamiltonian as was pointed out by Fano \cite{fano1981stark}.
However, in the case of non hydrogenic atoms in presence of external fields the barrier effects are significant especially at negative energies.  Therefore the use of solely the principal value Green's function $ G^{(C+F)}_P$ would not allow a straightforward implementation of scattering boundary conditions.
This is why the second term in Eq.~(\ref{greenpv}) has been included.

From the equality between Eqs.~(\ref{eq8}) [or (\ref{smoothspher})] and (\ref{eq9}), hereafter with the additional use of Eq.~(\ref{eq4}), the mapping of the irregular solutions is given by the following expression:
\begin{equation}
 \frac{g_{\epsilon \ell m}(\mathbf{r})}{r}=\sum_{n_1} \chi_{n_1 m}^{\epsilon F}(\mathbf{r})\csc(\gamma_{n_1})(\underline{U})^{\epsilon F m}_{n_1 \ell }~~\rm{for}~~ r \ll F^{-1/2}.
 \label{eq10}
\end{equation}
 
Additionally, Eq.~(\ref{eq4}) conventionally can be written as 
\begin{equation}
 \frac{f_{\epsilon \ell m}(\mathbf{r})}{r} =\sum_{n_1} \psi_{n_1 m}^{\epsilon F}(\mathbf{r})\big[(\underline{U}^{T})^{-1}\big]_{n_1 \ell}^{\epsilon Fm}, ~~\rm{for}~~ r \ll F^{-1/2},
 \label{eq11a}
\end{equation}
Note that in Eqs.~(\ref{eq10}) and (\ref{eq11a})  $\underline{U}^T$ and $[\underline{U}^T]^{-1}$ are the transpose and inverse transpose matrices of the $U$ LFT matrix whose elements are given by $(\underline{U})^{\epsilon F m}_{n_1 \ell}= U_{n_1 \ell}^{\epsilon Fm}$.

In Ref.\cite{stevens1996precision} Stevens {\it et al.} comment that in Eq.~(\ref{eq10}) only the left hand side possesses a uniform shift over the $\theta$-angles.
Quantifying this argument, one can examine the difference the semiclassical phases with and without the electric field.
Indeed, for a zero energy electron the phase accumulation due to the existence of the electric field as a function of the angle $\theta$ obeys the expression
\begin{eqnarray}
\Delta \phi(r,~\theta)&=& \int^r k(r,~\theta) dr-\int^r k_0(r)dr \nonumber \\
&\approx&-\frac{\sqrt{2}}{5} F r^{5/2} \cos\theta,~~\rm{for}~~Fr^2\ll1,
\label{approx}
\end{eqnarray}
where $k(r,~\theta)$ ($k_0(r)$) indicates the local momentum with (without) the electric field $F$.
In Eq.~(\ref{approx}) it is observed that for field strength $F=1$ kV/cm and $r<50$ a.u. the phase modification due to existence of the electric field is less than 0.001 radians.
This simply means that at short distances both sides of Eq.~(\ref{eq10}) should exhibit practically uniform phase over the angle $\theta$.

Recapitulating Eqs.~(\ref{eq10}) and (\ref{eq11a}) constitute the mapping of the regular and irregular functions respectively from spherical to parabolic cylindrical coordinates.

\section{Scattering observables in terms of the local frame transformation}
This section implements Harmin frame transformation theory to determine all the relevant scattering observables.

\subsection{The asymptotic form of the frame transformed irregular solution and the reaction matrix}
The irregular solutions which we defined in Eq.~(\ref{irregU}) are not the {\it usual} ones of the scattering theory since in the asymptotic region, namely $\eta \to \infty$, they do not lag by $\pi/2$ the regular functions, Eq.~(\ref{regU}).
Hence, this particular set of irregular solutions should not be used in order to obtain the scattering observables which are properly defined in the asymptotic region.

However, by linearly combining Eqs.~(\ref{regU}) and (\ref{irregU}) we define a new set of irregular solutions which are energy-normalized, asymptotically lag by $\pi/2$ the regular ones, and read:
\begin{equation}
 \bar{\Upsilon}_{n_1 m}^{\epsilon F,~\rm{scat}}(\eta)= \frac{1}{\sin \gamma_{n_1}}\bar{\Upsilon}_{n_1 m}^{\epsilon F}(\eta)-\cot \gamma_{n_1} \Upsilon_{n_1 m}^{\epsilon F}(\eta),
 \label{eq11}
\end{equation}
where this equation together with Eq.~(\ref{regU}) correspond to a set of real irregular and regular solutions according to the usual conventions of scattering theory.

The derivation of the reaction matrix follows.
Eqs.~(\ref{eq11}) and (\ref{eq7}) are combined and then substituted into Eq.~(\ref{eq10}) such that the irregular solution in spherical coordinates is expressed in terms of the $\bar{\Upsilon}_{n_1 m}^{\epsilon F,~\rm{scat}}$.
\begin{equation}
 \frac{g_{\epsilon \ell m}(\mathbf{r})}{r}=\sum_{n_1} \big[\psi_{n_1 m}^{\epsilon F}(\mathbf{r})\cot(\gamma_{n_1})+\chi_{n_1 m}^{\epsilon F,~\rm{scat}}(\mathbf{r})\big](\underline{U}^T)^{\epsilon F m}_{\ell n_1},
\label{eq12}
 \end{equation}
where $\chi_{n_1 m}^{\epsilon F,~\rm{scat}}(\mathbf{r})$ defined as 
\begin{equation}
 \chi_{n_1 m}^{\epsilon F,~\rm{scat}}(\mathbf{r})=e^{i m \phi} \Xi_{n_1 m}^{\epsilon F}(\xi) \bar{\Upsilon}_{n_1 m}^{\epsilon F,~\rm{scat}}(\eta)/\sqrt{2 \pi \xi \eta}.
\end{equation}

Hereafter, the short-range wave function ( Eq.~(\ref{spherwave})) expressed in spherical coordinates is transformed via the LFT $U$ into the asymptotic wave function.
Specifically,
\small
\begin{eqnarray}
\Psi_{\epsilon \ell m} (\mathbf{r})&=&\sum_{n_1} \psi_{n_1 m}^{\epsilon F}(\mathbf{r}) \bigg[  \big[(\underline{U}^{T})^{-1}\big]^{\epsilon F m}_{n_1 \ell} \cos \delta_\ell -  \cot \gamma_{n_1} (\underline{U})^{\epsilon F m}_{n_1 \ell}\times \nonumber \\
&\times& \sin \delta_\ell\bigg]-\chi_{n_1 m}^{\epsilon F}(\mathbf{r}) (\underline{U})^{\epsilon F m}_{ n_1 \ell } \sin \delta_\ell,
\label{eq13}
\end{eqnarray}
\normalsize

Then from  Eq.~(\ref{eq13}) and after some algebraic manipulations the reaction matrix solutions are written in a compact matrix notation as
\begin{eqnarray}
 \mathbf{\Phi}^{(R)}(\mathbf{r})&=& \mathbf{\Psi}[\cos \underline{\delta}]^{-1}\underline{U}^T[I-\cot \underline{\gamma} \underline{U}\tan\underline{\delta}\underline{U}^T ]^{-1} \\ \nonumber
 &=&\bar{\psi}(\mathbf{r})-  \bar{\chi}(\mathbf{r})[\underline{U}\tan \underline{\delta} \underline{U}^T][I-\cot \underline{\gamma} \underline{U}\tan\underline{\delta}\underline{U}^T ]^{-1},
 \label{eq14}
\end{eqnarray}
where $I$ is the identity matrix, the matrices $\cos \underline{\delta},~~\tan \underline{\delta}$, and $\cot \underline{\gamma}$ are diagonal ones.
Note that $\bar{\psi}$ ($\bar{\chi}$) indicates a vector whose elements are the $\psi_{n_1 m}^{\epsilon F}(\mathbf{r})$ ($\chi_{n_1 m}^{\epsilon F}(\mathbf{r})$) functions.
Similarly, the elements of the vector $\mathbf{\Psi}$ are provided by Eq.~(\ref{spherwave}).
Then from Eq.~(\ref{eq14}) the reaction matrix obeys the following relation:
\begin{equation}
 \underline{R}=\underline{U}\tan \underline{\delta}~\underline{U}^T \bigg[I- \cot \underline{\gamma}\underline{U}\tan \underline{\delta}~\underline{U}^T \bigg]^{-1},
 \label{eq15}
\end{equation}
In fact the matrix product $\underline{U}~\tan \underline{\delta}~\underline{U}^T$ can be viewed as a reaction matrix $\mathcal{\underline{K}}$ which does not encapsulates the impact of the Stark barrier on the wave function.
Moreover, as shown in Ref.\cite{robicheaux97} the recasting of the expression for the reaction matrix $\underline{R}$ in form that does not involve the inverse $[\underline{U}^T]^{-1}$ improves its numerically stability.
In addition, it can be shown with simple algebraic manipulations that the reaction matrix is symmetric. 
Note that this reaction matrix $R$ should not be confused with the Wigner-Eisenbud R-matrix.

The corresponding {\it physical} $S$-matrix is defined from the $R$-matrix via a Cayley transformation, namely
\begin{eqnarray}
  \underline{S}&=&\bigg[I+i \underline{R}\bigg]\bigg[I-i \underline{R}\bigg]^{-1} \nonumber \\
 &=&\bigg[I-\big(\cot \underline{\gamma}-i I\big)\mathcal{\underline{K}}\bigg]\bigg[I-\big(\cot \underline{\gamma}+iI\big)\mathcal{\underline{K}}\bigg]^{-1},
  \label{eq16}
\end{eqnarray}
where clearly this $S$-matrix is equivalent to the corresponding result of Ref.\cite{zhao12}.
Also, the $S$-matrix in Eq.~(\ref{eq16}) is unitary since the corresponding $R$-matrix is real and symmetric.

\subsection{Dipole matrix and outgoing wave function with the atom-radiation field interaction}

As was already discussed, the pair of parabolic regular and irregular solutions $\{\psi,\chi\}$ are the standing-wave solutions of the corresponding Schr\"odinger equation.
However, by linearly combining them and using Eq.~(\ref{eq14}), the corresponding energy-normalized outgoing/incoming wave functions are expressed as:
\begin{eqnarray}
\tilde{\mathbf{\Psi}}^{\pm}(\mathbf{r})&=& \mp \mathbf{\Phi}_{R}(\mathbf{r})\big[I\mp i \underline{R}\big]^{-1}\nonumber \\
&=&\frac{\mathbf{X}^{\mp}(\mathbf{r})}{i\sqrt{2}}-\frac{\mathbf{X}^{\pm}(\mathbf{r})}{i\sqrt{2}}\bigg[I\pm i \underline{R}\bigg]\bigg[I\mp i \underline{R}\bigg]^{-1},
\label{eq17}
\end{eqnarray}
where the elements of the vectors $\mathbf{X}^{\pm}\mathbf{r}$ are defined by the relation $[\mathbf{X}^\pm(\mathbf{r})]^{\epsilon F}_{n_1 m}=(-\chi_{n_1 m}^{\epsilon F} (\mathbf{r})\pm i \psi_{n_1 m}^{\epsilon F} (\mathbf{r}))/\sqrt{2}$.

In the treatment of the photoionization of alkali-metal atoms, the dipole matrix elements are needed to compute the cross sections which characterize the excitation of the atoms by photon absorption.
Therefore, initially we assume that at small distances the short-range dipole matrix elements possess the form $d_\ell=\bra{\Psi_{\epsilon \ell m}}\hat{\varepsilon} \cdot \hat{r} \ket{\Psi_{\rm{init}}}$.
Note that the term $\hat{\varepsilon} \cdot \hat{r}$ is the dipole operator, the $\hat{\varepsilon}$ denotes the polarization vector and $\ket{\Psi_{\rm{init}}}$ indicates the initial state of the atom.
Then the dipole matrix elements which describe the transition amplitudes from the initial to each $n_1$-th of the reaction-matrix states is 
\begin{equation}
D_{n_1}^{(R)}=\sum_\ell d_\ell \big\{[\cos\underline{\delta}]^{-1} \underline{U}^T\big[I-\cot \underline{\gamma}\mathcal{\underline{K}}\big]^{-1}\big\}_{\ell n_1}.
\label{eq18}
 \end{equation}

Now with the help of Eq.~(\ref{eq18}) we define the dipole matrix elements for transitions from the initial state to the incoming wave final state which has only outgoing waves in the $\mathbf{n_1-th}$ channel.
The resulting expression is
\begin{equation}
 D^{(-)}_{n_1}=\sum_{n^{\prime}_1}D_{n^{\prime}_1}^{(R)} \big[(I-i \underline{R})^{-1}\big]_{n^{\prime}_1 n_1}. 
\label{eq19}
 \end{equation}

Eq.~(\ref{eq19}) provides the necessary means to properly define the outgoing wave function with the atom-field radiation.
As it was shown in Ref.~\cite{zhao12b} the outgoing wave function can be derived as a solution of an inhomogeneous Schr\"odinger equation that describes the atom being perturbed by the radiation field.
Formally this implies that
\begin{equation}
 [\epsilon-H]\Psi_{\rm{out}}(\mathbf{r})= \hat{\varepsilon} \cdot \hat{r} \Psi_{\rm{init}}(\mathbf{r}),
 \label{20}
\end{equation}
where $\Psi_{\rm{out}}(\mathbf{r})$ describes the motion of the electron after its photoionization moving in the presence of an electric filed,  $H$ is the Stark Hamiltonian with $\epsilon$ being the energy of ionized electron.
The $\Psi_{\rm{out}}(\mathbf{r})$ can be expanded in outgoing wave functions involving the dipole matrix elements of Eq.~(\ref{eq19}).
More specifically we have that
\begin{equation}
 \Psi_{\rm{out}}(\mathbf{r})=\sum_{n_1 m} D^{(-)}_{n_1 m}X_{n_1 m}^{\epsilon F,~+}(\mathbf{r}).
 \label{eq21}
\end{equation}

\subsection{Wave function microscopy and differential cross sections} 
Recent experimental advances \cite{cohen13,itatani2004tomographic,bordas03,Nicole02} have managed to detect the square module of the electronic wave function, which complements a number of corresponding theoretical proposals \cite{ost1,ost2,ost3,ost4}.
This has been achieved by using a position-sensitive detector to measure the flux of slow electrons that are ionized in the presence of an electric field.  

The following defines the relevant observables associated with the photoionization-microscopy.
The key quantity is the differential cross section which in turn is defined through the electron current density.
As in Ref.~\cite{zhao12b}, consider a detector placed beneath the atomic source with its plane being perpendicularly to the axis of the electric field.
Then, with the help of Eq.~(\ref{eq21}) the electron current density in cylindrical coordinates has the following form:
\begin{equation}
 R(\rho, z_{\rm{det}},\phi)=\frac{2 \pi \omega}{c}\rm{{\it Im}} \big[-\Psi_{out}(\mathbf{r})^\ast\frac{d}{dz}\Psi_{out}(\mathbf{r})\bigg]_{z=z_{\rm{det}}},
 \label{eq22}
\end{equation}
where $z_{\rm{det}}$ indicates the position of the detector along the $z$-axis, $c$ is the speed of light and $\omega$ denotes the frequency of the photon being absorbed by the electron.
The integration of the azimuthal $\phi$ angle leads to the differential cross section per unit length in the $\rho$ coordinate.
Namely, we have that
\begin{equation}
 \frac{d \sigma(\rho, z_{\rm{det}})}{d\rho}=\int_0^{2 \pi}d\phi~\rho R(\rho, z_{\rm{det}},\phi),
 \label{eq23}
\end{equation}

\section{Eigenchannel R-matrix calculation}

Harmin's Stark effect theory for nonhydrogenic atoms is mainly based on the semi-classical WKB approach.
In order to eliminate the WKB approximation as a potential source of error, this section implements a fully quantal description of Harmin's theory based on a variational eigenchannel R-matrix calculation  as was formulated in Ref.\cite{Greene1983pra,GreeneKim1988pra} and reviewed in \cite{aymar1996multichannel}.  As implemented here using a B-spline basis set, the technique also shares some similarities with the Lagrange-mesh R-matrix formulation developed by Baye and coworkers\cite{baye2010}.
The present application to a 1D system with both an inner and an outer reaction surface accurately determines regular and irregular solutions of the Schr\"odinger equation in the $\eta$ degrees of freedom. The present implementation can be used to derive two independent solutions of any one-dimensional Schr\"odinger equation of the form
 \begin{equation}
\bigg[-\frac{1}{2} \frac{d^2}{d\eta^2} + V(\eta) \bigg]\psi(\eta)= \frac{1}{4} \epsilon \psi(\eta),
\label{eq25}
\end{equation}
where 
\begin{equation}
V(\eta)=\frac{m^2-1}{8 \eta^2}-\frac{1-\beta}{2\eta}-\frac{F}{8}\eta
\label{eq26}
\end{equation}

The present application of the non-iterative eigenchannel R-matrix theory adopts a reaction surface $\Sigma$ with two disconnected parts, one at an inner radius $\eta_1$ and the other at an outer radius $\eta_2$.  The reaction volume $\Omega$ is the region $\eta_1 < \eta < \eta_2$.

This one-dimensional R-matrix calculation is based on the previously derived variational principle \cite{FanoLeePRL,Greene1983pra} for the eigenvalues $b$ of the R-matrix,
\begin{equation}
 b[\psi]=\frac{\int_\Omega\big[- \overrightarrow{\nabla}\psi^\ast \cdot \overrightarrow{\nabla}\psi+2\psi^\ast (E-V) \psi \big]d \Omega}{\int_\Sigma \psi^\ast \psi d\Sigma}.
 \label{eq27}
\end{equation}

Physically, these R-matrix eigenstates have the same outward normal logarithmic derivative everywhere on the reaction surface consisting here of these two points $\Sigma_1$ and $\Sigma_2$.  
The desired eigenstates obey the following boundary condition:
\begin{equation}
 \frac{\partial \psi}{\partial n}+b \psi=0,~~\rm{on}~\Sigma .
\label{eq26b}
\end{equation}

In the present application the $\psi$-wave functions are expanded as a linear combination of a nonorthogonal B-spline basis \cite{deBoor}, i.e. 

\begin{equation}
 \psi(\eta)=\sum_{i} P_i B_i(\eta) = \sum_{C} P_C B_C(\eta)+ P_IB_I(\eta)+ P_OB_O(\eta),
 \label{eq28}
\end{equation}
where $P_i$ denote the unknown expansion coefficients and $B_i(\eta)$ stands for the B-spline basis functions.
The first term in the left hand side of Eq.~(\ref{eq28}) was regarded as the ``closed-type basis set” in \cite{aymar1996multichannel} because every function $B_c(\eta)$ vanishes on the reaction surface, i.e. $B_c(\eta_1)= B_c(\eta_2)=0$. 
The two basis functions $B_I(\eta)$ and $B_O(\eta)$ correspond to the ``open-type basis functions’’ of Ref. \cite{aymar1996multichannel} in that they are the only B-spline functions that are nonzero on the reaction surface.
Specifically, only $B_I(\eta)$ is nonzero on the inner surface $\eta=\eta_1$ ($\Sigma_1$) and only $B_O(\eta)$ is nonzero on the outer surface $\eta=\eta_2$ ($\Sigma_2$). 
Moreover the basis functions $B_I$ and $B_O$ have no region of overlap in the matrix elements discussed below.

Insertion of this trial function into the variational principle leads to the following generalized eigenvalue equation:

\begin{equation}
\underline{\Gamma} P=b \underline{\Lambda} P.
\label{eq29}
\end{equation}

In addition, the real, symmetric matrices $\underline{\Gamma}$ and $\underline{\Lambda}$ are given by the following expressions for this one-dimensional problem:
\small
\begin{eqnarray}
 \Gamma_{i j}&=&\int_{\eta_1}^{\eta_2} \big[(\frac{1}{2} \epsilon -2 V(\eta) ) B_i(\eta)B_j(\eta) + B_i^{\prime}(\eta)B_j^{\prime}(\eta) \big]d \eta, \\
 \Lambda_{ij}&=& B_i(\eta_1) B_j(\eta_1)+ B_i(\eta_2) B_j(\eta_2)=\delta_{i,I}\delta_{I,j} + \delta_{i,O}\delta_{O,j}, 
\end{eqnarray}
\normalsize
where $\delta$ indicates the Kronecker symbol and the $^{\prime}$ are regarded as the derivatives with respect to the $\eta$.

It is convenient to write this linear system of equations in a partitioned matrix notation, namely:
\begin{eqnarray}
\underline{\Gamma}_{CC}P_C +\underline{\Gamma}_{CI}P_I+ \underline{\Gamma}_{CO}P_O &=&0 \\
\underline{\Gamma}_{IC}P_C+ \underline{\Gamma}_{II}P_I &=&b P_I \\
\underline{\Gamma}_{OC}P_C+ \underline{\Gamma}_{OO}P_O &=&b P_O. 
\end{eqnarray}
Now the first of these three equations is employed to eliminate $P_C$ by writing it as $P_C = -\underline{\Gamma}_{CC}^{-1}\underline{\Gamma}_{CI}P_I -\underline{\Gamma}_{CC}^{-1}\underline{\Gamma}_{CO}P_O$, which is equivalent to the ``streamlined transformation'' in Ref.\cite{GreeneKim1988pra}.
This gives finally a 2$\times$2 matrix $\underline{\Omega}$ to diagonalize at each $\epsilon$ in order to find the two R-matrix eigenvalues $b_\lambda$ and corresponding eigenvectors $P_{i\lambda}$: 
\begin{eqnarray}
\bigg(\begin{matrix}
	  \underline{\Omega}_{II} & \underline{\Omega}_{IO} \\[0.3em]
	  \underline{\Omega}_{OI} & \underline{\Omega}_{OO} \\[0.3em]
      \end{matrix}
\bigg)
 \bigg(\begin{matrix}
	  P_{I}    \\[0.3em]
	  P_{O}  \\[0.3em]
        \end{matrix}
\bigg)
= b
\bigg(\begin{matrix}
	P_{I}    \\[0.3em]
        P_{O}  \\[0.3em]
      \end{matrix}
\bigg).
\label{eq32}
\end{eqnarray}
Here, e.g., the matrix element $\underline{\Omega}_{II} \equiv \underline{\Gamma}_{II}-\underline{\Gamma}_{IC}\underline{\Gamma}_{CC}^{-1}\underline{\Gamma}_{CI}$, etc.

In any 1D problem like the present one, the use of a B-spline basis set leads to a banded structure for $\underline{\Gamma}_{CC}$ which makes the construction of $\underline{\Gamma}_{CC}^{-1} \underline{\Gamma}_{CI}$ and $\underline{\Gamma}_{CC}^{-1} \underline{\Gamma}_{CO}$ highly efficient in terms of memory and computer processing time;  this step is the slowest in this method of solving the differential equation, but still manageable even in complex problems where the dimension of $\underline{\Gamma}_{CC}$ can grow as large as 10$^4$ to $10^5$.

Again, the indices $C$ refer to the part of the basis expansion that is confined fully within the reaction volume and vanishes on both reaction surfaces.

The diagonalization of Eq.~(\ref{eq32}) provides us with the $b_\lambda$--eigenvalues and the corresponding eigenvectors, which define two linearly independent wave functions $\psi_\lambda$, with $\lambda=1,2$.
These obey the Schr\"odinger equation, Eq.~\ref{eq25} and have equal normal logarithmic derivatives at $\eta_1$ and $\eta_2$. 
The final step is to construct two linearly independent solutions that coincide at small $\eta$ with the regular and irregular field-free $\eta$-solutions $f_{\epsilon \beta m}(\eta)$ and $g_{\epsilon \beta m}(\eta)$ (Cf. Appendix \ref{app:coulomhalfint}). 
These steps are rather straightforward and are not detailed further in this paper.

\section{Results and discussion}

\subsection{The frame transformed irregular solutions}

\begin{figure}[t!]
\includegraphics[scale=0.58]{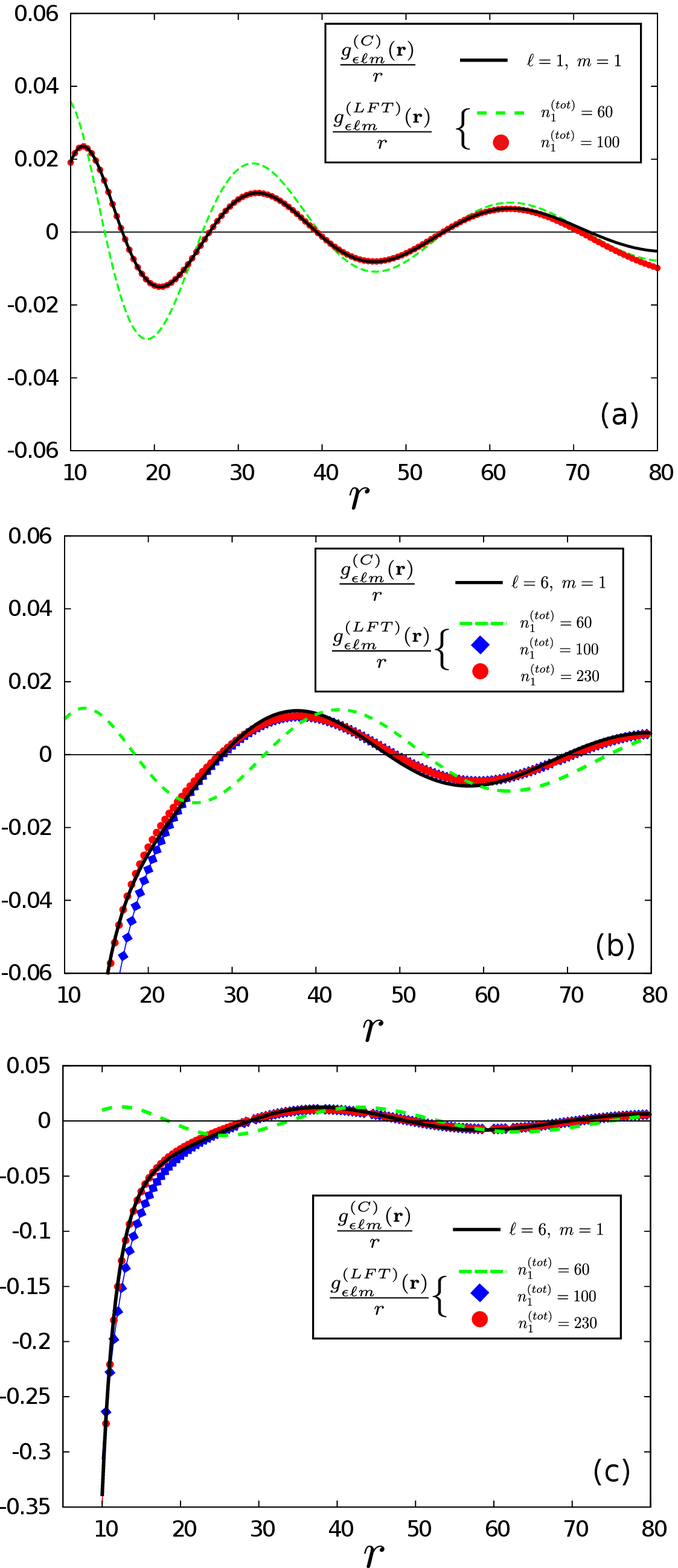}
\caption{(color online). The irregular solutions in spherical coordinates illustrated up to $r=80$ au where $\mathbf{r}=(r,~\theta=5\pi/6,~\phi=0)$. In all panels the azimuthal quantum number is set to $m=1$ and the black solid line indicates the irregular coulomb function in spherical coordinates, namely $\frac{g_{\epsilon \ell m}^{(C)}(\mathbf{r})}{r}$. (a) depicts the case of $\ell=1$ where $\frac{g_{\epsilon \ell m}^{(LFT)}(\mathbf{r})}{r}$ denotes the irregular function in spherical coordinates calculated within the local frame transformation (LFT) framework, for two different cases of total amount of $n_1$ states, namely $n_1^{(tot)}=60$ (green dashed line) and $n_1^{(tot)}=100$ (red dots). (b) refers to the case of $\ell=6$ where $\frac{g_{\epsilon \ell m}^{(LFT)}(\mathbf{r})}{r}$ is calculated for $n_1^{(tot)}=60$ (green dashed line), $n_1^{(tot)}=100$ (blue diamonds) and $n_1^{(tot)}=230$ (red dots) states. Note that panel (c) is a zoomed-out plot of the curves shown in panel (b).}
\label{fig2}
\end{figure}

To reiterate, Zhao et al. \cite{zhao12} claim that the Fano-Harmin frame transformation is inaccurate, based on a disagreement between their full numerical calculations of the differential cross section and the LFT calculation.  They then claim to have investigated the origin of the discrepancy and pinpointed an error in the frame transformed irregular function. The present section carefully tests the main conclusion of Ref. \cite{zhao12} that Eq.~(\ref{eq10}) does not accurately yield the development of the irregular spherical Coulomb functions into a parabolic field-dependent solution (see Fig.5 in Ref. \cite{zhao12}). 

Fig.\ref{fig2} illustrates the irregular solutions in spherical coordinates where $\mathbf{r}=(r,~\theta=5\pi/6,~\phi=0)$ and the azimuthal quantum number is set to be $m=1$. 
The energy is taken to be $\epsilon=135.8231$ cm$^{-1}$ and the field strength is $F=640$ V/cm.
In addition we focus on the regime where $r<~90~ \rm{au} \ll F^{-1/2}$.
In all the panels the black solid line indicates the analytically known irregular Coulomb function, namely $\frac{g_{\epsilon \ell m}^{(C)}(\mathbf{r})}{r}$. 
Fig.\ref{fig2}(a) and (b,c) examine the cases of angular momentum $\ell=1$ and $6$, respectively.
All the green dashed lines, the diamonds and dots correspond to the frame transformed irregular Coulomb functions in spherical coordinates, namely  $\frac{g_{\epsilon \ell m}^{(LFT)}(\mathbf{r})}{r}$, which are calculated by summing up from 0 to $n_1$ the irregular $\chi_{n_1 m}^{\epsilon F}(\mathbf{r})$ functions in the parabolic coordinates as Eq.~(\ref{eq10}) indicates.

\begin{figure}[t!]
\includegraphics[scale=0.5]{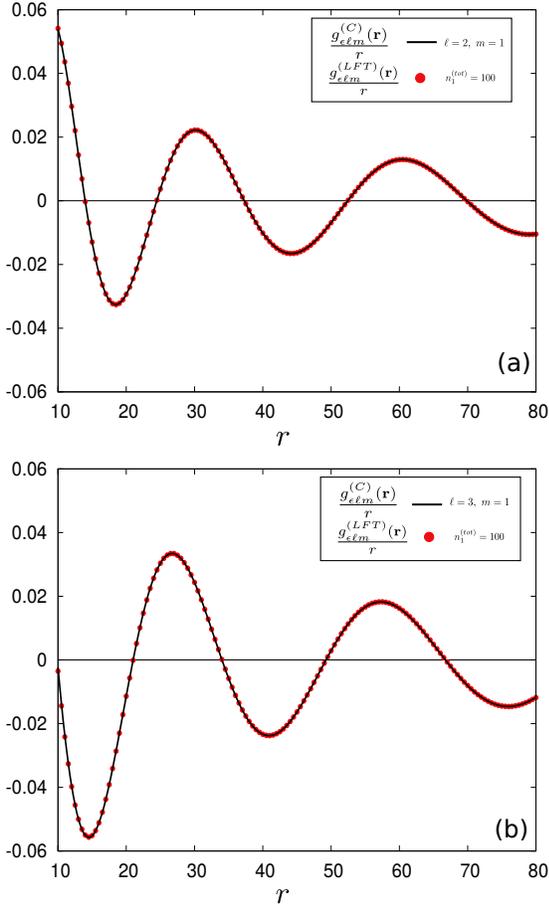}
\caption{(color online). The irregular solutions in spherical coordinates are shown up to $r=80$ au where $\mathbf{r}=(r,~\theta=5\pi/6,~\phi=0)$. In all panels the azimuthal quantum number is set to be $m=1$ and the black solid line indicates the analytically known irregular coulomb function, namely $\frac{g_{\epsilon \ell m}^{(C)}(\mathbf{r})}{r}$. (a) depicts the case of $\ell=2$ where $\frac{g_{\epsilon \ell m}^{(LFT)}(\mathbf{r})}{r}$ denotes the irregular function in spherical coordinates calculated within the local frame transformation (LFT) framework for $n_1^{(tot)}=100$ states (red dots). Similarly, (b) refers to the case of $\ell=3$ with $\frac{g_{\epsilon \ell m}^{(C)}(\mathbf{r})}{r}$ being calculated for $n_1^{(tot)}=100$ (red dots) states.}
\label{fig3}
\end{figure}

The positive value of the energy ensures that all the $n_1$-channels lie well above the local maximum in the $\eta$ whereby the phase parameter $\gamma_{n_1}$ is very close to its semiclassical expected value $\pi/2$.
Furthermore, since only short distances are relevant to this comparison, namely $r<~90$ au, this means that the summed $\bar{\Upsilon}_{n_1 m}^{\epsilon F}(\eta)$ functions on the right-hand side of Eq.~(\ref{eq10}) in the $n_1$-th irregular $\chi_{n_1 m}^{\epsilon F}(\mathbf{r})$ will be equal to analytically known Coulomb irregular functions in the parabolic coordinates.  
This is justified since at the interparticle distances that we are interested in, namely $\ll F^{-1/2}$, the electric field is negligible in comparison to the Coulomb interaction. 
Then the corresponding Schr\"odinger equation becomes equal to the Schr\"odinger equation of the pure Coulomb field which is analytically solvable in spherical and parabolic coordinates as well. 
Thus, in the following we employ the above-mentioned considerations in the evaluation of the right hand side of Eq.~(\ref{eq10}) for Figs.\ref{fig2} and \ref{fig3}.

\begin{figure}[t!]
\includegraphics[scale=0.45]{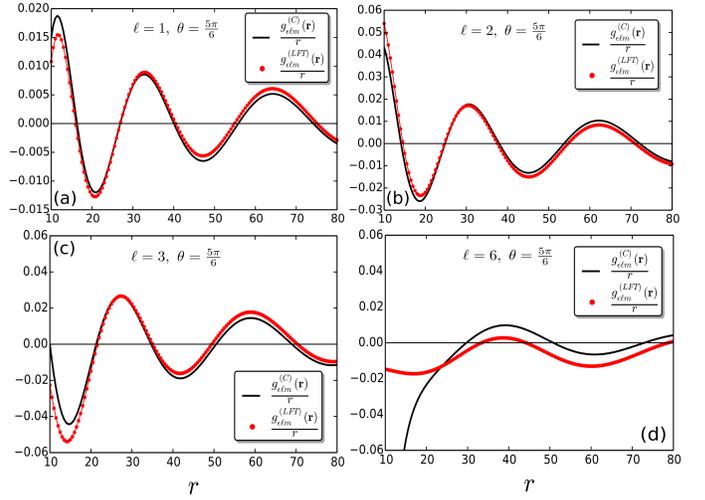}
\caption{(color online). The irregular solutions in spherical coordinates at negative energies, ie $\epsilon=-135.8231$ cm$^{-1}$, illustrated for $\mathbf{r}=(r,~\theta= \frac{5 \pi}{6},~\phi=0)$. In all panels the azimuthal quantum number is set to be $m=1$ and the black solid line indicates the analytically known irregular coulomb function, namely $\frac{g_{\epsilon \ell m}^{(C)}(\mathbf{r})}{r}$. Accordingly, the red dots correspond to the LFT calculations of irregular function, namely $\frac{g_{\epsilon \ell m}^{(LFT)}(\mathbf{r})}{r}$. Panels (a-d) depict the cases of $\ell=1,~2,~3~\rm{and}~6$, respectively. For all the LFT calculations the total amount of $n_1$ states is $n_1^{(tot)}=25$ which corresponds to $\beta_{n_1}<1$.}
\label{fig4}
\end{figure}

Fig.\ref{fig2}(a) compares the radial irregular Coulomb function (black line) with those calculated in the LFT theory for $\ell=m=1$.
In order to check the convergence of the LFT calculations with respect to the total number $n_1^{(\rm{tot})}$ different values are considered. 
Indeed, we observe that the $\frac{g_{\epsilon \ell m}^{(LFT)}(\mathbf{r})}{r}$ for $n_1^{(tot)}=60$ (green dashed line) does not coincide with $\frac{g_{\epsilon \ell m}^{(C)}(\mathbf{r})}{r}$ (black line) particularly in the interval of small interparticle distances $r$.  
This can be explained with the help of Fig.\ref{fig1}, which demonstrates that the LFT $U$ for $\ell=1$ possesses nonzero elements for $n_1> 60$, and those elements are crucial for the growth of the irregular solution at small distances.
Therefore, the summation in Eq.~\ref{eq10} for $\ell=1$ does not begin to achieve convergence until $n_1\ge 100$, where the corresponding elements of the LFT $U$ tend to zero.
Indeed, when the sum over $n_1$ states is extended to this larger range, the irregular function $\frac{g_{\epsilon \ell m}^{(LFT)}}{r}$ of LFT theory, i.e. for $n_1^{(tot)}=100$ (red dots), accurately matches the spherical field-free irregular solution $\frac{g_{\epsilon \ell m}^{(C)}(\mathbf{r})}{r}$ (black line) (see Fig.\ref{fig2}) at small electron distances $r$.

\begin{figure*}[t!]
\includegraphics[scale=0.7]{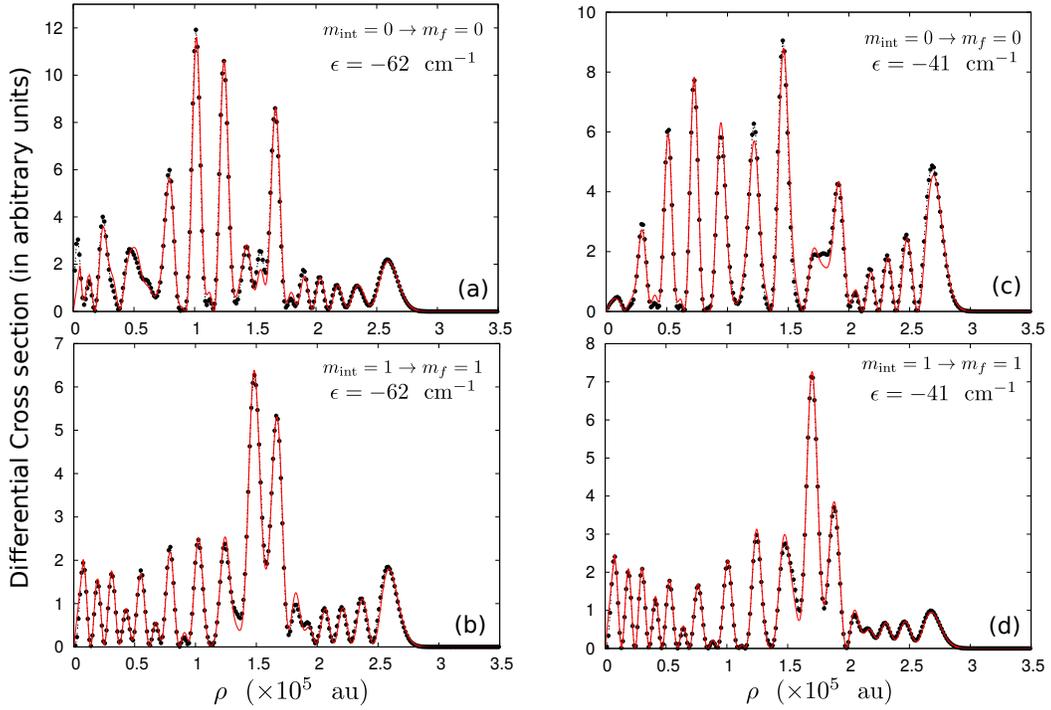}
\caption{(color online). The differential cross section for Na atoms as a function of the cylindrical coordinate $\rho$. The red solid lines indicate the LFT theory calculations, whereas the black dots denote the velocity mapping results from a direct solution of the two-dimensional inhomogeneous Schr\"odinger equation. Panels (a) and (b) they refer to energy $\epsilon=-62$ cm$^{-1}$ for the transitions $m_{\rm{int}}=0\rightarrow m_f=0$ and $m_{\rm{int}}=1\rightarrow m_f=1$, respectively. Similarly, panels (c) and (d) they refer to energy $\epsilon=-41$ cm$^{-1}$ for the transitions $m_{\rm{int}}=0\rightarrow m_f=0$ and $m_{\rm{int}}=1\rightarrow m_f=1$, respectively.
In all cases the field strength is $F=3590$ V/cm and the detector is placed at $z_{\det}=-1$~mm.}
\label{fig5}
\end{figure*}

Furthermore, Fig.(\ref{fig2})(b) refers to the case of $\ell=6~m=1$. 
Specifically, for $n_1^{tot}=60$ states the $\frac{g_{\epsilon \ell m}^{(LFT)}}{r}$ (green dashed line) agrees poorly with the $\frac{g_{\epsilon \ell m}^{(C)}}{r}$ (black line).
Though as in the case of $\ell=1$, by increasing the number of $n_1$ states summed over in Eq.~(\ref{eq10}) the corresponding $\frac{g_{\epsilon \ell m}^{(LFT)}}{r}$, namely to $n_1^{(tot)}=100$ (blue diamonds) and to  $n_1^{(tot)}=230$ (red dots), better agreement is achieved with the $\frac{g_{\epsilon \ell m}^{(C)}}{r}$.
In contrast to the case where $\ell=1$, the convergence is observed to be very slow for $\ell=6$. The main reason for this is that for $r<20$ au we are in the classically forbidden region where $\frac{g_{\epsilon \ell m}^{(LFT)}}{r}$ diverges as $1/r^{\ell+1}$.
From Eq.~(\ref{eq10}) it is clear that the sum will diverge due to the divergent behavior of the irregular functions of the $\eta$ direction, namely the $\bar{\Upsilon}_{n_1 m}^{\epsilon F}(\eta)$.
Hence, in order the $\bar{\Upsilon}_{n_1 m}^{\epsilon F}(\eta)$ to be divergent in the interval of 10 to 80 au it is important to take into account many $n_1$ states which correspond to $\beta_{n_1}>1$ since only then the term $1-\beta/\eta$ becomes repulsive and producing the diverging behavior appropriate to a classical forbidden region.
Fig.\ref{fig2}(c) is a zoomed-out version of the functions shown in panel (b), which demonstrates that the $\frac{g_{\epsilon \ell m}^{(LFT)}}{r}$ for $n_1^{(tot)}=230$ correctly captures the divergent behavior of $\frac{g_{\epsilon \ell m}^{(C)}}{r}$ for $r<20$.

Similarly, Fig.\ref{fig3} explores the cases of $\ell=2$ (see Fig.\ref{fig3}(a)) and $\ell=3$ (see Fig.\ref{fig3}(b)).
In both panels the black solid lines indicate the field free Coulomb function in spherical coordinates $\frac{g_{\epsilon \ell m}^{(C)}}{r}$ and the red dots correspond to the $\frac{g_{\epsilon \ell m}^{(LFT)}}{r}$ for $n_1^{(tot)}=100$.
Both panels exhibit $\frac{g_{\epsilon \ell m}^{(LFT)}}{r}$ that are in excellent agreement with $\frac{g_{\epsilon \ell m}^{(C)}}{r}$.

Having analyzed the LFT calculations at positive energies, Fig. \ref{fig4} illustrates the corresponding LFT calculations at negative energies, namely $\epsilon=-135.8231$ cm$^{-1}$ where the field strength is set to be $F=640~~\rm{V/cm}$.
Note that these parameters \cite{note} are used for an analogous comparison in Fig.5 of Ref. \cite{zhao12}.
In all panels the azimuthal quantum number is considered to be $m=1$, the solid black lines denote the analytically known irregular Coulomb function $[\frac{g_{\epsilon \ell m}^{(C)}(\mathbf{r})}{r}]$ and red dots refer to the corresponding LFT calculations $[\frac{g_{\epsilon \ell m}^{(LFT)}(\mathbf{r})}{r}]$.
In Fig. \ref{fig4}(a-d) the $\ell=1,~2,~3~\rm{and}~6$ cases are considered at $\mathbf{r}=(r,~\theta= \frac{5 \pi}{6},~\phi=0)$, respectively.
In addition, for all the panels of Fig. \ref{fig4} in the LFT calculations the summation over the $n_1$ states is truncated at $n_1^{\rm{tot}}=25$ for the considered energy and field strength values.
This simply means that in the summation of the framed-transformed irregular function contribute solely all the fractional charges $\beta_{n_1}$ that obey the relation $\beta_{n_1}<1$.
These states essentially describe all the relevant physics since only for these states the ``down field'' part of the wave function can probe the core either above or below the Stark barrier.
Therefore, the $n_1$ states for which $\beta_{n_1}>1$ physically are irrelevant since they yield a strongly repulsive barrier in the ``down field'' degree of freedom shielding completely the core.
However, for these states the considered pair of regular and irregular functions in Sec. II C for the $\eta$-degree of freedom acquire imaginary parts due to the fact that the colliding energy is below the minimum of the corresponding Coulomb potential.
Consequently, these states are omitted from the sum of the frame-transformed irregular function.
The omission of states with $\beta_{n_1}>1$ mainly addresses the origin of the accuracy in the LFT calculations.

The impact of the omitted states is demonstrated in Fig. \ref{fig4} where discrepancies are observed as the orbital angular momentum $\ell$ increases since more $n_1$ states are needed.
Indeed, in panels (a), (b) and (c) of Fig. \ref{fig4} a good agreement is observed between the framed-transformed irregular function and the Coulombic one (black solid line). 
On the other hand, in panel (d) of Fig. \ref{fig4} small discrepancies, particularly for $r>20$ are observed occurring due to poor convergence over the summation of the $n_1$ states.
Though, these discrepancies are of minor importance since they correspond to negligible quantum defects yielding thus minor contributions in the photoabsorption cross section.

The bottom line of the computations shown in this subsection is that the frame-transformed irregular functions $\frac{g_{\epsilon \ell m}^{(LFT)}}{r}$ do not display, at least for $\ell= 1~\rm{or}~2$, the inaccuracies that were claimed by Zhao {\it et al.} in Ref.\cite{zhao12}.
For negative energies, our evidence suggests that the inclusion of $n_1$ states with $\beta_{n_1}>1$ will enhance the accuracy of the frame-transformed irregular functions as it is already demonstrated by the LFT calculations at positive energies.

\subsection{Photoionization microscopy}
Next we compute the photoionization microscopy observable for Na atoms, namely the differential cross section in terms of the LFT theory.
The system considered is a two step photoionization of ground-state Na in the presence of an electric field $F$ of strength 3590 V/cm, which is again the same system and field strength treated in Ref.\cite{zhao12}.
The two consecutive laser pulses are assumed to be $\pi$ polarized along the field axis, which trigger in succession the following two transitions: (i) the excitation of the ground state to the intermediate state $^2 P_{3/2}$, namely $[\rm{Ne}]~3s~~^2S_{1/2}\rightarrow[\rm{Ne}]~3p~~^2P_{3/2}$ and (ii) the ionization from the intermediate state $^2 P_{3/2}$.
In addition, due to spin-orbit coupling the intermediate state will be in a superposition of the states which are associated with different orbital azimuthal quantum numbers, i.e. $m=0$ and $1$. 
Hyperfine depolarization effects are neglected in the present calculations.

Fig. \ref{fig5} illustrates the differential cross section $\frac{d\sigma(\rho, z_{\rm{det}})}{d \rho}$ for Na atoms, where the detector is placed at $z_{\rm{det}}=-1$~mm and its plane is perpendicular to the direction of the electric field.
Since spin-orbit coupling causes the photoelectron to possess both azimuthal orbital quantum numbers $m=0,1$, the contributions from both quantum numbers are explored in the following. Fig.\ref{fig5} panels (a) and (c) illustrate the partial differential cross section for transitions of $m_{\rm{int}}=0 \rightarrow m_f=0$, where $m_{\rm{int}}$ indicates the {\it intermediate} state azimuthal quantum number and $m_f$ denotes the corresponding quantum number in the final state.
Similarly, panels (b) and (d) in Fig.\ref{fig5} are for the transitions $m_{\rm{int}}=1 \rightarrow m_f=1$.
In addition, in all panels of Fig.\ref{fig5} the red solid lines correspond to the LFT calculations, whereas the black dots indicate the {\it ab initio} numerical solution of the inhomogeneous Schr\"odinger equation which employ a velocity mapping technique and which do not make use of the LFT  approximation.

More specifically, this method uses a discretization of the Schr\"odinger equation on a grid of points in the radial coordinate $r$ and an orbital angular momentum grid in $\ell$. 
The main framework of the method is described in detail in Sec.~2.1 of Ref.~\cite{TR1} and below only three slight differences are highlighted.
In order to represent a cw-laser, the source term was changed to $S_0(\vec{r},t)=[1+\rm{erf}(t/t_w)]z\psi_{init}(\vec{r})$ with $\psi_{\rm{init}}$ either the $3p,~m=0$ or $3p,~m=1$ state. 
The time dependence, $1+\rm{erf}(t/t_w)$ gives a smooth turn-on for the laser with time width of $t_w$; $t_w$ is chosen to be of the order a few picoseconds. 
The second difference is that the Schr\"odinger equation is solved until the transients from the laser turn on decayed to zero. 
The last difference was in how the differential cross section is extracted. 
The radial distribution in space slowly evolves with increasing distance from the atoms and the calculations become challenging as the region represented by the wave function increases. 
To achieve convergence in a smaller spatial region,  the velocity distribution in the $\rho$-direction is directly obtained.
The wave function in $r,~\ell$ is numerically summed over the orbital angular momenta $ \ell$ yielding $\psi_m(\rho,z)$ where $m$ is the azimuthal angular momentum. 
Finally, using standard numerical techniques a Hankel transformation is performed on the wave function $\psi_m(\rho,z)$ which reads 
\begin{equation}
\psi_m(k_\rho,z)=\int d\rho \rho J_m(k_\rho\rho)\psi_m(\rho,z)
\end{equation}
which can be related to the differential cross section.
The cross section is proportional to $k_\rho |\psi_m(k_\rho,z)|^2$ in the limit that $z\to -\infty$. 
The $k_\rho$ is related to the $\rho$ in Fig.\ref{fig5} through a scaling factor. 
The convergence of our results is tested with respect to number of angular momenta, number of radial grid points, time step, $|z|_{max}$, $t_w$ and final time.
The bandwidth that the following calculations exhibit is equal to $0.17$ cm$^{-1}$.
In addition, in order to check the validity of our velocity mapping calculation we directly compute numerically the differential cross section through the electron flux defined in Eq.~(\ref{eq22}).
An agreement of the order of percent is observed solidifying our investigations.

One sees immediately in panels (a-d) of Fig.\ref{fig5}  that the LFT calculations are in good agreement with the full numerical ones, with only minor areas of disagreement.
In particular, the interference patterns in all calculations are essentially identical.
An important point is that panels (a) and (c) do not exhibit the serious claimed inaccuracies of the LFT approximation that were observed in Ref.\cite{zhao12}. 
In fact, the present LFT calculations are in excellent agreement with the corresponding LFT calculations of Zhao {\it et al.}
Evidently, this suggests that the disagreement observed by the Zhao {\it et al} originates from coupled-channel calculations and not the LFT theory, in particular for the case of $m=0$.
Indeed, panels (b) and (d) of Fig.\ref{fig5} are in excellent agreement with the corresponding results of both the LFT and coupled-channel calculations of Ref.\cite{zhao12}.

\section{Summary and conclusions}
The present study reviews Harmin's Stark-effect theory and develops a standardized form of the corresponding LFT theory.
In addition, the LFT Stark-effect theory is formulated in the traditional framework of scattering theory including its connections to the photoionization observables involving the dipole matrix elements, in particular the differential cross section.
In order to quantitatively test the LFT, the present formulation does not use semi-classical WKB theory as was utilized by Harmin.  Instead the one-dimensional differential equations are solved within an eigenchannel R-matrix framework. 
This study has thoroughly investigated the core idea of the LFT theory, which in a nutshell defines a mapping between the irregular solutions of two regions, namely spherical solutions in the field-free region close to the origin and the parabolic coordinate solutions relevant from the core region all the way out to asymptotic distances.
For positive energies, our calculations demonstrate that indeed the mapping formula Eq.~(\ref{eq10}) predicts the correct Coulomb irregular solution in spherical coordinates (see Figs.\ref{fig2} and \ref{fig3}).
On the other hand, at negative energies it is demonstrated (see Fig.\ref{fig4}) that the summation over solely ``down field'' states $\beta_{n_1}<1$ imposes minor limitations in the accuracy of LFT calculation mainly for $\ell>3$.
Our study also investigates the concept of wave function microscopy through calculations of photoionization differential cross sections for a Na atom in the presence of a uniform electric field. 
The photoionization process studied is a resonant two-photon process where the laser field is assumed to be $\pi$ polarized.
The excellent agreement between the LFT and the full velocity mapping calculation has been conclusively demonstrated, and the large discrepancies claimed by Ref.\cite{zhao12} in the case of $m_{\rm{int}}=0 \rightarrow m_f=0$ are not confirmed by our calculations.

These findings suggest that the LFT theory passes the stringent tests of wave function microscopy, and can be relied upon both to provide powerful physical insight and quantitatively accurate observables, even for a complicated observable such as the differential photoionization cross section in the atomic Stark effect.    

\begin{acknowledgements}
The authors acknowledge Ilya Fabrikant and Jesus Perez-Rios for helpful discussions.
The authors acknowledge support from the U.S. Department of Energy Office of Science, Office of Basic Energy Sciences Chemical Sciences, Geosciences, and Biosciences Division under Award Numbers DE-SC0012193 and DE-SC0010545.
\end{acknowledgements}

\appendix
\section{Coulomb functions for non-positive half-integer angular momentum at negative energies}
\label{app:coulomhalfint}
In this appendix we will present the regular and irregular Coulomb functions with non-positive half-integer, either positive or negative, quantum numbers.
The necessity for this particular type of solutions arises from the fact that they constitute the boundary conditions for the R-matrix eigenchannel calculations in the 'down field' $\eta$ degree of freedom at sufficient small distances where essentially the field term can be neglected. 
This corresponds in the field free case where the orbital angular momentum does not possess non-negative integer values.

The Schr\"odinger equation in the field free case for the $\eta$ parabolic coordinate has the following form
\begin{equation}
  \frac{d^2}{d\eta^2}f_{\beta m}^{ \epsilon }(\eta)+\bigg(\frac{\epsilon}{2}+\frac{1-m^2}{4\eta^2}+\frac{1-\beta}{\eta} \bigg)f_{\beta m}^{ \epsilon} (\eta)=0,
 \label{A1}
\end{equation}
where the energy $\epsilon$ is considered to be negative.
Assuming that $\bar{\epsilon}=2 \epsilon/(1-\beta)^2$, $\zeta=\frac{1-\beta}{2} \eta$ and $\lambda=(m-1)/2$ Eq.~(\ref{A1}) can be transformed into the following differential equation:
\begin{equation}
   \frac{d^2}{d\zeta^2}f_{\lambda}^{ \bar{\epsilon} }(\zeta)+\bigg(\bar{\epsilon}-\frac{\lambda(\lambda+1)}{\zeta^2}+\frac{2}{\zeta} \bigg)f_{\lambda}^{ \bar{\epsilon}} (\zeta)=0,
 \label{A2}
\end{equation}
which for integer $\lambda$ has two linearly independent energy normalized solutions whose relative phase is $\pi/2$ at small distances and negative energies
\small
\begin{eqnarray}
 f_{\lambda}^{\bar{\epsilon}} (\zeta) &=& A(\bar{\nu}, \lambda)^{1/2}S_{\lambda}^{\bar{\epsilon}}(\zeta) \label{a3a} \\
 g_{\lambda}^{\bar{\epsilon}} (\zeta) &=& A(\bar{\nu}, \lambda)^{1/2}S_{\lambda}^{\bar{\epsilon}}(\zeta) \cot((2 \lambda+1)\pi)\nonumber \\
 &-&\frac{A(\bar{\nu}, \lambda)^{-1/2}S_{-\lambda-1}^{\bar{\epsilon}}(\zeta)}{\sin((2 \lambda+1)\pi)},
 \label{a3}
 \end{eqnarray}
\normalsize
where $\bar{\nu}=1/\sqrt{-\bar{\epsilon}}$, $A(\bar{\nu}, \lambda)=\frac{\Gamma(\lambda+\bar{\nu}+1)}{\bar{\nu}^{2\lambda+1}\Gamma(\bar{\nu}-\lambda)}$ and the function $S_{\lambda}^{\bar{\epsilon}}(\zeta)$ is obtained by the following relation
\begin{equation}
 S_{\lambda}^{\bar{\epsilon}}(\zeta)=2^{\lambda+1/2}\zeta^{\lambda+1}e^{-\zeta/\bar{\nu}}~_1\bar{F}_1(\lambda-\bar{\nu}+1;2+2\lambda;2 \zeta/\bar{\nu}),
 \label{a4}
\end{equation}
where the function $_1\bar{F}_1$ denotes the regularized hypergeometric function $_1F_1$.
One basic property of this function is that it remains finite even when its second argument is a non-positive integer. 
We recall that the hypergeometric $_1F_1(a;b;x)$ diverges when $b=-1, -2,-3,..$. 

Moreover, we observe that when $\lambda$ acquires half-integer values, ie $\lambda=\lambda_c$ the nominator and denominator of $g_{\lambda}^{\bar{\epsilon}}$ in Eq.~(\ref{a3}) both vanish.
Therefore, employing the de l' Hospital's theorem on $g_{\lambda}^{\bar{\epsilon}}$ in Eq.~(\ref{a3}) we obtain the following expression:
\small
\begin{eqnarray}
  \bar{g}_{\lambda_c}^{\bar{\epsilon}} (\zeta)&=& \frac{1}{2 \pi}\frac{\partial f_{\lambda}^{\bar{\epsilon}} (\zeta)}{\partial \lambda}\Bigg|_{\lambda=\lambda_c}-\frac{1 }{2\pi \cos[(2\lambda_c+1)\pi]}\frac{\partial f_{-\lambda-1}^{\bar{\epsilon}} (\zeta)}{\partial \lambda}\Bigg|_{\lambda=\lambda_c}. \label{a5}
\end{eqnarray}
\normalsize

Hence, Eqs. (\ref{a3a}) and (\ref{a5}) correspond to the regular and irregular Coulomb functions for non-positive half-integers at negative energies, respectively.
This particular set of solutions possess $\pi/2$-relative phase at short distances and they used as boundary conditions in the eigenchannel R-matrix calculations.
A similar construction is possible with the help of Ref. \cite{olver2010nist} for positive energies but it is straightforward and not presented here.

\bibliography{qdtbiblio}

\begin{thebibliography}{43}
\bibitem [{not()}]{note}%
  \BibitemOpen
  \href@noop {} {}\bibinfo {note} {In Fig. 5 of Ref. [29] the energy is
  $\epsilon=-135.8231$ cm$^{-1}$ (personal communication).}\BibitemShut {Stop}%
\end{thebibliography}

\end{document}